\documentclass[aps,prb,twocolumn,10pt,groupedaddress,noeprint]{revtex4-2}

\usepackage{graphicx}
\usepackage{amsfonts}
\usepackage{amsmath}
\usepackage{amssymb}
\usepackage{mathtools}
\usepackage{siunitx}
\usepackage{enumerate}
\usepackage{cases}
\usepackage{bm}
\usepackage{empheq}
\usepackage{multirow}
\usepackage{physics}
\usepackage[pdfusetitle,colorlinks,allcolors=blue,pdftex,
pdfauthor={Matteo Mariantoni},
pdftitle={Interacting Defects Generate Stochastic Fluctuations in Superconducting Qubits},
pdfkeywords={Quantum Computing, Superconducting Resonators, Superconducting Qubits, Amorphous Dielectric Materials, Two-Level Systems, Standard Tunneling Model, Stochastic Fluctuations, Spectral Diffusion, Generalized Tunneling Model, Allan Deviation}]{hyperref}

\begin{document}

\title{Interacting Defects Generate Stochastic Fluctuations in Superconducting Qubits}

\author{J. H.~B\'{e}janin}
\affiliation{Institute for Quantum Computing, University of Waterloo, 200 University Avenue West, Waterloo, Ontario N2L 3G1, Canada}
\affiliation{Department of Physics and Astronomy, University of Waterloo, 200 University Avenue West, Waterloo, Ontario N2L 3G1, Canada}

\author{C. T.~Earnest}
\affiliation{Institute for Quantum Computing, University of Waterloo, 200 University Avenue West, Waterloo, Ontario N2L 3G1, Canada}
\affiliation{Department of Physics and Astronomy, University of Waterloo, 200 University Avenue West, Waterloo, Ontario N2L 3G1, Canada}

\author{A. S.~Sharafeldin}
\thanks{Present address: Department of Engineering Physics, McMaster University, Hamilton, Ontario L8S 4L7, Canada.}
\affiliation{Institute for Quantum Computing, University of Waterloo, 200 University Avenue West, Waterloo, Ontario N2L 3G1, Canada}
\affiliation{Department of Physics and Astronomy, University of Waterloo, 200 University Avenue West, Waterloo, Ontario N2L 3G1, Canada}

\author{M.~Mariantoni}
\email[Corresponding author: ]{matteo.mariantoni@uwaterloo.ca}
\affiliation{Institute for Quantum Computing, University of Waterloo, 200 University Avenue West, Waterloo, Ontario N2L 3G1, Canada}
\affiliation{Department of Physics and Astronomy, University of Waterloo, 200 University Avenue West, Waterloo, Ontario N2L 3G1, Canada}

\date{\today}

\begin{abstract}
Amorphous dielectric materials have been known to host two-level systems~(TLSs) for more than four decades. Recent developments on superconducting resonators and qubits enable detailed studies on the physics of TLSs. In particular, measuring the loss of a device over long time periods (a few days) allows us to investigate stochastic fluctuations due to the interaction between TLSs. We measure the energy relaxation time of a frequency-tunable planar superconducting qubit over time and frequency. The experiments show a variety of stochastic patterns that we are able to explain by means of extensive simulations. The model used in our simulations assumes a qubit interacting with high-frequency TLSs, which, in turn, interact with thermally activated low-frequency TLSs. Our simulations match the experiments and suggest the density of low-frequency TLSs is about three orders of magnitude larger than that of high-frequency ones.
\end{abstract}


\maketitle

\section{INTRODUCTION}
	\label{Sec:INTRODUCTION}

Superconducting devices operated in the quantum regime~\cite{Kjaergaard:2020} are ideal tools to study the properties of amorphous dielectric materials~\cite{Mueller:2019}. These materials are known to be characterized by defects that can be modeled as two-level systems~(TLSs)~\cite{Phillips:1987}. TLSs can interact with superconducting resonators or qubits, resulting in dissipation channels that are particularly prominent in planar devices. Such devices are fabricated by depositing superconducting films made from metals, e.g., aluminum~(Al) or niobium, on silicon~(Si) or sapphire substrates. A few examples of planar devices can be found in our works of Refs.~\cite{Earnest:2018} and \cite{Bejanin:2020}, where we have investigated coplanar waveguide~(CPW) resonators~\cite{Frunzio:2005} as well as Xmon transmon qubits~\cite{Barends:2013}.

A large body of work on CPW resonators and qubits has shown that TLSs are likely hosted in native oxide layers~\cite{Martinis:2005,Gao:2008:b,Wisbey:2010,Sage:2011,Megrant:2012,Richardson:2016,Dunsworth:2017,DeGraaf:2018,Moeed:2019,Bilmes:2021} at the substrate-metal~(SM), substrate-air~(SA), or metal-air~(MA) interfaces~\cite{Wenner:2011:b,Gambetta:2017:a,Earnest:2018,Woods:2019}. TLSs originate within these layers because naturally occurring oxides deviate from crystalline order. This deviation may result in trapped charges, dangling bonds, tunneling atoms, or collective motion of molecules.

It is convenient to distinguish between two categories of~TLSs based on their energy~$E$ and the device operating temperature~$T$. When~$E > k_{\text{B}} T$, the corresponding~TLSs reside in the quantum ground state; these TLSs are hereafter referred to as quantum-TLSs~(Q-TLSs). When~$E < k_{\text{B}} T$, the TLSs are thermally activated and are referred to as thermal-TLSs~(T-TLSs). Typically, superconducting resonators are characterized by a resonance frequency~$f_{\text{r}}$ and qubits by a transition frequency~$f_{\text{q}}$, with~$f_{\text{r}} \sim f_{\text{q}} \sim \SI{5}{\giga\hertz}$, and are operated at~$T \sim \SI{50}{\milli\kelvin}$. Hence, the energy threshold between Q- and T-TLSs is~$E/h \sim \SI{1}{\giga\hertz}$.

Superconducting quantum devices interact \mbox{(semi-)}resonantly with Q-TLSs~\cite{Neeley:2008}, affecting the internal quality factor of resonators, $Q_{\text{i}}$, or the energy relaxation time of qubits, $T_1$. Several authors have hypothesized that Q-TLSs additionally interact with T-TLSs~\cite{Burnett:2014,Faoro:2015,Mueller:2015}, leading to experimentally observed stochastic fluctuations in~$Q_{\text{i}}$ and $f_{\text{r}}$~\cite{Neill:2013,Burnett:2014,DeGraaf:2018,Moeed:2019} as well as~$T_1$ and $f_{\text{q}}$~\cite{Paik:2011,Mueller:2019}. The model proposed by these authors depart from the TLS standard tunneling model~(STM), where TLS interactions are neglected~\cite{Phillips:1987}. The interacting model is sometimes called the \emph{generalized tunneling model}~(GTM).

It has recently been shown that planar fixed-frequency transmon qubits exhibit random fluctuations in both~$T_1$ and $f_{\text{q}}$ over very long time periods~\cite{Burnett:2019,Schloer:2019}. Frequency-tunable transmon qubits, as the Xmon, show TLS-induced fluctuations predominantly in~$T_1$~\cite{FrequencyFluctuations,Klimov:2018}. These findings serve as the main motivation for the experiments and simulations presented in this article.

In this article, we present the experimental measurement of spectrotemporal charts for an Xmon transmon qubit as well as the results of detailed simulations corresponding to these experiments. In the spectrotemporal charts, $T_1$ is measured and simulated for time periods up to~\SI{48}{\hour} and for~$f_{\text{q}}$ ranges up to~\SI{300}{\mega\hertz}. Our main objective is to validate the Q-TLS--T-TLS interaction hypothesis in the GTM by comparing experiments and simulations. In our simulations, a qubit interacts with an ensemble of Q-TLSs, the frequencies of which undergo stochastic fluctuations due to the interaction with T-TLSs. For every Q-TLS we consider a set of interacting T-TLSs, where the dynamics of each T-TLS state are governed by a random telegraph signal~(RTS). The Q-TLS frequency fluctuation process, which is broadly referred to as \emph{spectral diffusion}, is responsible for the random fluctuations in~$T_1$.

The comparison between experiments and simulations reveals that the Q-TLS--T-TLS interaction likely exists, as proposed in the GTM. In particular, our simulations reproduce well the spectral-diffusion patterns presented in the experiments. Our model suggests that the density of T-TLSs is significantly higher than that of Q-TLSs. We find a T-TLS density of approximately~\SI{6e+05}{\per\giga\hertz\per\micro\meter\cubed}, which is about three orders of magnitude larger than the Q-TLS density.

Finally, we show that certain statistical analyzes, such as the Allan deviation, are not able to capture the fluctuation characteristics of a given time series (e.g., the number of T-TLSs contributing to the stochastic process). Instead, a direct analysis of the time series provides a more accurate description of the stochastic processes due to TLSs.

The article is organized as follows. In Sec.~\ref{Sec:THEORY}, we review the theory necessary to describe the stochastic fluctuations of~$T_1$. In Sec.~\ref{Sec:METHODS}, we explain the methods required to perform experiments and simulations. In Sec.~\ref{Sec:RESULTS}, we present our main results. In Sec.~\ref{Sec:DISCUSSION}, we provide an in-depth discussion on some of our main results. Finally, in Sec.~\ref{Sec:CONCLUSIONS}, we summarize our findings and suggest a roadmap for future work.

\section{THEORY}
	\label{Sec:THEORY}

In this section, we introduce physical models of TLSs (Subsec.~\ref{Subsec:Physical:Models:of:TLSs}); we then describe the qubit--Q-TLS and Q-TLS--T-TLS interaction (Subsecs.~\ref{Subsec:Qubit--Q-TLS:Interaction} and \ref{Subsec:Q-TLS--T-TLS:Interaction}); finally, we amalgamate the previous concepts in order to explain qubit stochastic fluctuations (Subsec.~\ref{Subsec:Qubit:Stochastic:Fluctuations}).

\subsection{Physical Models of TLSs}
	\label{Subsec:Physical:Models:of:TLSs}

The STM is a phenomenological model describing defects in amorphous dielectric materials. The defects are commonly assumed to be quantum-mechanical double-well potentials, or TLSs, with energy barrier~$V$. In the STM, the TLS tunneling energy~$\Delta_0$ is calculated by means of the WKB approximation,
\begin{equation}
\Delta_0 \simeq h \Omega_0 \exp \left( - \dfrac{d}{\hbar} \sqrt{2 \, m \, V} \right) .
	\label{Eq:Delta0}
\end{equation}
In this equation, $\Omega_0$ is the attempt frequency (assumed to be the same for both wells), $d$ is the spatial distance between the two wells, and~$m$ is the mass of the physical entity associated with the TLS (e.g., a molecular mass)~\cite{Esquinazi:1998}.

The unperturbed Hamiltonian of a TLS reads~$\widehat{H}_{\text{TLS}} = \left( \Delta \hat{\bar{\sigma}}_z + \Delta_0 \hat{\bar{\sigma}}_x \right) / 2$, where~$\Delta$ is the asymmetry energy between the two wells of the TLS; $\hat{\bar{\sigma}}_z$ and $\hat{\bar{\sigma}}_x$ are the usual Pauli matrices in the so-called diabatic (``left'' and ``right'') basis. By diagonalizing this Hamiltonian we obtain~$\widehat{H}_{\text{TLS}} = E \, \hat{\sigma}_z / 2$, where
\begin{equation}
E = \sqrt{\Delta^2 + \Delta^2_0}
	\label{Eq:E}
\end{equation}
is the TLS energy and $\hat{\sigma}_z = [ \hat{\bar{\sigma}}_z \, \cos(\theta) + \hat{\bar{\sigma}}_x \, \sin(\theta) ] / 2$ is the Pauli matrix in the energy basis; $\theta = \arctan(\Delta_0/\Delta)$ is the rotation angle used to perform the diagonalization.

One of the hypothesis in the STM is that~$\Delta$ and $\Delta_0$ are uncorrelated quantities with joint probability density
\begin{align}
	f_{\Delta , \Delta_0} =
		\begin{cases}
			\dfrac{D}{\Delta_0} ,
			& \text{for} \enskip \Delta \geq 0
			\enskip \text{and} \enskip
			\Delta_0 \geq \mathcal{E}_{\text{min}} ;
\\[2.5mm]
			0 ,
			& \text{otherwise} .
		\end{cases}
	\label{Eq:fDeltaDelta0:STM}
\end{align}
In this equation, $D$ is the TLS density in units of inverse energy and volume and $\mathcal{E}_{\text{min}}$ is the minimum tunneling energy. A further hypothesis is that interactions between TLSs are very weak and, thus, \emph{negligible}.

The hypotheses behind the STM prevent this model from explaining a variety of features observed in devices affected by TLS defects. Among other phenomena, the STM cannot explain the temperature dependence of the frequency noise of superconducting resonators~\cite{Burnett:2014} as well as the strong temperature dependence of the relaxation rate of Q-TLSs measured with qubits~\cite{Lisenfeld:2010}. Most importantly, the STM cannot explain the spectral diffusion dynamics observed both in the work of Ref.~\cite{Klimov:2018} and in our experiments.

In order to resolve these shortcomings, it is necessary to extend the STM to the GTM by making the following modifications:
\begin{enumerate}[(1)]
\item Interactions between TLSs are not neglected.
\item The joint probability density is assumed to be nonuniform with respect to~$\Delta$,
\begin{align}
	f_{\Delta , \Delta_0} =
		\begin{cases}
			\dfrac{1 + \mu}{\Delta_0} \left( \dfrac{\Delta}{\mathcal{E}_\text{max}} \right)^{\!\! \mu} ,
			& \text{for} \enskip 0 \leq \Delta \leq \mathcal{E}_{\text{max}}
			\\[-1.5mm]
			&
			\text{and} \enskip \mathcal{E}_{\text{min}} \leq \Delta_0 \leq \mathcal{E}_{\text{max}} ;
\\[5.0mm]
			0 ,
			& \text{otherwise} .
		\end{cases}
	\label{Eq:fDeltaDelta0:GTM}
	\raisetag{2.5mm}
\end{align}
In this equation, $\mu < 1$ is a small positive parameter and $\mathcal{E}_{\text{max}}$ is a maximum energy cutoff dictated by the energy scales of the system under consideration (see Subsec.~\ref{Subsec:METHODS:Simulations}).
\end{enumerate}

The interaction energy between any pairs of TLSs is assumed to be a function of their spatial separation~$r$,
\begin{equation}
U(r) = \dfrac{U_0}{r^3} ,
	\label{Eq:Ur}
\end{equation}
where~$U_0$ is a material-dependent parameter associated with electric or elastic interactions. It is worth noting that interactions can occur between pairs of Q-TLSs or T-TLSs as well as between a T-TLS and a Q-TLS.

In the study of superconducting planar qubits, both~$f_{\text{q}}$ and $T_1$ are affected by the interactions hypothesized in the GTM. These type of qubits interact semi-resonantly with an ensemble of Q-TLSs, where each Q-TLS can strongly interact with one or more T-TLSs. Such interactions lead to stochastic fluctuations in~$T_1$ and $f_{\text{q}}$.

\subsection{Qubit--Q-TLS Interaction}
	\label{Subsec:Qubit--Q-TLS:Interaction}

The interaction between a qubit and a single Q-TLS leads to perturbations in~$T_1$ and $f_{\text{q}}$. These perturbations depend on the coupling strength between the qubit and Q-TLS, $g$, and on the difference between the Q-TLS transition frequency~$f_{\text{Q-TLS}}$ and $f_{\text{q}}$, $\Delta f = f_{\text{q}} - f_{\text{Q-TLS}}$. In this work, we consider only~$T_1$ fluctuations because, for a tunable qubit, $f_{\text{q}}$ fluctuations are dominated by other noise processes such as flux noise.

In the rotating frame of the qubit and after a rotating wave approximation, the Hamiltonian of the qubit coupled to the Q-TLS reads
\begin{equation}
\widehat{H}_{\text{q,Q-TLS}} =
	h \Delta f \, \hat{\sigma}_{\text{q}}^{+} \hat{\sigma}_{\text{q}}^{-} + h g \left( \hat{\sigma}_{\text{q}}^{+} \otimes \hat{\sigma}_{\text{Q-TLS}}^{-} + \text{H.c.} \right) ,
	\label{Eq:H_qQ-TLS}
\end{equation}
where~$\hat{\sigma}_{\text{q}}^{\mp}$ and $\hat{\sigma}_{\text{Q-TLS}}^{\mp}$ are the qubit and Q-TLS lowering and raising operators in the energy basis and H.c.~is the Hermitian conjugate of the first term in parentheses. The coupling strength~$g$ is due to the electric dipole moment~$\vec{p}$ of the Q-TLS and the electric field~$\vec{E}_{\text{q}}$ of the qubit~\footnote{The electric field~$\vec{E}_{\text{q}}$ is the field associated with the qubit capacitor, which is described in Subsec.~\ref{Subsec:METHODS:Experiments} and App.~\ref{App:QUBIT:ELECTRIC:FIELD}.}, $h g = \vec{p} \cdot \vec{E}_{\text{q}}$.

The contribution to the energy relaxation rate of the qubit due to the Q-TLS can be approximated by
\begin{equation}
\Gamma_1^{\text{q,Q-TLS}} = \dfrac{\Gamma_1^{\text{Q-TLS}} - \widetilde{\Gamma}_1^{\text{q}} - \Re [\Lambda]}{2} ,
	\label{Eq:Gamma1qQ-TLS}
\end{equation}
where~$\Gamma_1^{\text{Q-TLS}}$ is the energy relaxation rate of the Q-TLS due to phononic interactions with the environment, $\widetilde{\Gamma}_1^{\text{q}}$ is the bare energy relaxation rate of the qubit~\footnote{This is the rate caused by all dissipation sources other than TLSs.}, and
\begin{equation}
\Lambda = \sqrt{ \left( \widetilde{\Gamma}_1^{\text{q}} + 2 i (2 \pi \Delta f) - \Gamma_1^{\text{Q-TLS}} \right)^2 - 16 (2 \pi g)^2} ,
	\label{Eq:Lambda}
\end{equation}
with~$i^2 = -1$. Equation~(\ref{Eq:Gamma1qQ-TLS}) is valid when~$\Gamma_1^{\text{Q-TLS}} > \widetilde{\Gamma}_1^{\text{q}}$, which is typically the case in our devices. The derivation of Eq.~(\ref{Eq:Gamma1qQ-TLS}) is shown in App.~\ref{App:DERIVATION:OF:Gamma1qQTLS}.

In presence of amorphous dielectric materials, the qubit is coupled to an ensemble of Q-TLSs. In this case, Eq.~(\ref{Eq:Gamma1qQ-TLS}) represents the individual contribution to the energy relaxation rate of the qubit due to the~$k$-th Q-TLS, $\Gamma_1^{\text{q,Q-TLS}} \rightarrow \Gamma_1^{\text{q} , k}$; each Q-TLS is now characterized by its own coupling strength~$g_k$, frequency~$f_k$, and energy relaxation rate~$\Gamma_1^k$. The effective qubit relaxation rate is therefore given by
\begin{equation}
\Gamma_1^{\text{q}} = \dfrac{1}{T_1} = \widetilde{\Gamma}_1^{\text{q}} + \sum_k \Gamma_1^{\text{q} , k} .
	\label{Eq:Gamma1q}
\end{equation}

\subsection{Q-TLS--T-TLS Interaction}
	\label{Subsec:Q-TLS--T-TLS:Interaction}

We intend to calculate the frequency shift experienced by a Q-TLS due to the interaction with a T-TLS. We assume that the unperturbed energy and eigenstates are~$E = E_{\text{T-TLS}}$ and $\{ \ket{-} , \ket{+} \}$ for the T-TLS and $E = E_{\text{Q-TLS}} \gg E_{\text{T-TLS}}$ and $\{ \ket{0} , \ket{1} \}$ for the Q-TLS. These two TLSs form a quantum-mechanical system with Hamiltonian given by Eq.~(11) in the work of Ref.~\cite{Faoro:2015}. Assuming the interaction energy~$U$ between the T-TLS and Q-TLS is given by Eq.~(\ref{Eq:Ur}), the four eigenenergies of the system are
\begin{subequations}
	\begin{empheq}[]{align}
		E^{\mp}_0 & = -\dfrac{E_{\text{Q-TLS}}}{2} \mp \sqrt{\left( \dfrac{E_{\text{T-TLS}}}{2} \right)^2 + U \Delta + U^2}
		\label{SubEq:Emp0}
\\[2.0mm]
		\text{and} \nonumber
\\[2.0mm]
		E^{\mp}_1 & = +\dfrac{E_{\text{Q-TLS}}}{2} \mp \sqrt{\left( \dfrac{E_{\text{T-TLS}}}{2} \right)^2 - U \Delta + U^2} ,
		\label{SubEq:Emp1}
	\end{empheq}
\end{subequations}
where~$\Delta$ is the asymmetry energy of the T-TLS.

The frequency shift~$\delta \! f^{\mp}$ of the Q-TLS due to the interaction with the T-TLS reads
\begin{equation}
h \, \delta \! f^{\mp} = E^{\mp}_1 - E^{\mp}_0 - E_{\text{Q-TLS}} ,
	\label{Eq:hdeltafmp}
\end{equation}
which is negative when the T-TLS is in~$\ket{-}$ and positive otherwise.

A T-TLS is thermally activated because of the condition~$E_{\text{T-TLS}} < k_{\text{B}} T$ and, thus, switches state in time. This causes the sign of~$\delta \! f^{\mp}$ to change, affecting the time evolution of the frequency of the Q-TLS coupled to it.

\subsection{Qubit Stochastic Fluctuations}
	\label{Subsec:Qubit:Stochastic:Fluctuations}

We assume that the state of a T-TLS over time is modeled by an RTS with switching rate
\begin{equation}
\gamma = \gamma_0 \exp \left( - \dfrac{V}{k_{\text{B}} T} \right) ,
	\label{Eq:gamma}
\end{equation}
where~$\gamma_0$ is a heuristic proportionality constant and $V$ is implicitly given by Eq.~(\ref{Eq:Delta0}).

A Q-TLS is generally coupled to several T-TLSs, where the~$\ell$-th T-TLS is characterized by a certain value of~$\gamma_{\ell}$ and $\delta \! f^{\mp}_{\ell}$. Given the state ($\ket{\mp}$) of each T-TLS at a time~$t$, we can approximate the effective frequency shift of the Q-TLS by summing the individual values of~$\delta \! f^{\mp}_{\ell} (t)$. Since the T-TLS state is modeled by an RTS, the effective shift varies with~$t$ leading to a time series
\begin{equation}
f_{\text{Q-TLS}} (t) = \dfrac{E_{\text{Q-TLS}}}{h} + \sum_{\ell} \delta \! f^{\mp}_{\ell} (t) .
	\label{Eq:fQ-TLSt}
\end{equation}

For the $k$-th Q-TLS, $f_k$ fluctuates in time according to Eq.~(\ref{Eq:fQ-TLSt}). As a consequence, $\Gamma_1^{\text{q}}$ fluctuates because of its dependence on~$\Gamma_1^{\text{q} , k}$, which, in turn, depends on~$f_k$ through Eq.~(\ref{Eq:Lambda}). The stochastic fluctuations of~$\Gamma_1^{\text{q}} = 1 / T_1$ are the main subject of this article.

\section{METHODS}
	\label{Sec:METHODS}

In this section, we describe the methods used to perform the experiments on~$T_1$ fluctuations (Subsec.~\ref{Subsec:METHODS:Experiments}) and the corresponding simulations (Subsec.~\ref{Subsec:METHODS:Simulations}).

\subsection{Experiments}
	\label{Subsec:METHODS:Experiments}

In this work, we use an Xmon transmon qubit to probe TLS defects. The main goal of our experiments is to characterize fluctuations in~$T_1$ over long time periods and for different values of~$f_{\text{q}}$. We measure~$T_1$ by means of a standard energy relaxation experiment, a ``$T_1$ experiment.'' Details on the qubit and setup are given in App.~\ref{App:DEVICE:AND:SETUP}.

In a $T_1$ experiment, we prepare the qubit in the excited state~$\ket{\text{e}}$ by means of a~$\pi$ pulse. We then measure the average population of~$\ket{\text{e}}$, $P_{\text{e}}$, for many values of a delay time spaced logarithmically between~$\SI{1}{\nano\second}$ and $\SI{200}{\micro\second}$. Due to the various relaxation channels affecting the qubit, including TLS interactions, $P_{\text{e}}$ decays exponentially in time. We obtain~$T_1$ by fitting the exponential decay and acquire between~$36$ and $38$ points for each~$T_1$ experiment.

We measure~$T_1$ for different values of~$f_{\text{q}}$ by setting a quasi-static flux bias~$\phi_{Z}^{\text{qs}}$ applied to the qubit. The correspondence between~$\phi_{Z}^{\text{qs}}$ and $f_{\text{q}}$ is obtained from a qubit parameter calibration. Depending on the experiment, we set~$f_{\text{q}}$ over different bandwidths varying between~$30$ and $\SI{300}{\mega\hertz}$. We select~$N_f$ linearly spaced values of~$f_{\text{q}}$ for each~$T_1$ experiment. The~$T_1$ measurements are repeated continuously at a repetition period~$\Delta t$ over an observation time~$t_{\text{obs}}$, leading to matrices of data points as detailed in App.~\ref{App:EXPERIMENTAL:DETAILS}. These matrices constitute the spectrotemporal charts of~$T_1$ presented in Sec.~\ref{Sec:RESULTS}. The experimental parameters for the three datasets shown in this work are reported in Table~\ref{Tab:Experimental:Parameters}.

\begin{table}[b!]
	\caption{Experimental parameters for the three datasets introduced in Sec.~\ref{Sec:RESULTS}. Number of frequency points, $N_f$. Qubit frequency, $f_{\text{q}}$. Repetition period, $\Delta t$. Observation time, $t_{\text{obs}}$. \label{Tab:Experimental:Parameters}}
	\begin{center}
	\begin{ruledtabular}
		\begin{tabular}{ccccc}
			\raisebox{0mm}[4mm][0mm]{Dataset} & $N_f$ & $f_{\text{q}}$ range & $\Delta t$ & $t_{\text{obs}}$ \\
			\raisebox{0mm}[0mm][2mm]{} & {\footnotesize{$(-)$}} & \footnotesize{(\si{\giga\hertz})} & \footnotesize{(\si{\second})} & \footnotesize{(\si{\hour})} \\
			\hline
			\hline
			\\[-3.0mm]
			\raisebox{0mm}[3mm][0mm]{$1$} & $16$ & $[4.369,4.669]$ & $640$ & $42.5$ \\
			\hline
			\raisebox{0mm}[3mm][0mm]{$2$} & $31$ & $[4.500,4.560]$ & $1000$ & $47.2$ \\
			\hline
			\raisebox{0mm}[3mm][0mm]{$3$} & $31$ & $[4.500,4.530]$ & $1000$ & $48.1$ \\
		\end{tabular}
	\end{ruledtabular}
	\end{center}
\end{table}

\subsection{Simulations}
	\label{Subsec:METHODS:Simulations}

The procedure to simulate the effect of TLSs on the stochastic fluctuations in~$T_1$ is composed of three main steps: (1) Generate an ensemble of Q-TLSs interacting with the qubit. (2) Generate several T-TLSs interacting with each Q-TLS. (3) Generate a time series for each T-TLS and propagate the effect of the T-TLSs' switching state to each Q-TLS, and, finally, to the qubit.

Before detailing each step of the procedure, it is worth introducing a few general assumptions. We consider that all TLSs are distributed uniformly in the oxide layers at the SA and MA interfaces of the qubit device. The thickness of these layers is assumed to be~$t_{\text{ox}} = \SI{3}{\nano\meter}$ for both interfaces~\cite{Wenner:2011:b,Kamal:2016,McRae:2018}. We additionally set~$\widetilde{\Gamma}_1^{\text{q}} = \SI{1/27}{\mega\hertz}$. Finally, for all distributions used in this work, we determine the probability density function~(PDF) by normalizing a given distribution [e.g., that represented by Eq.~(\ref{Eq:fDeltaDelta0:GTM})] over the chosen boundary values; we also find the cumulative density function~(CDF). In order to pick a random value from a distribution, we generate a random quartile value between~$0$ and $1$. We then calculate the random value corresponding to the generated quartile either by inverting the CDF or via root finding.

For step (1), we follow a similar procedure as in the work of Ref.~\cite{Barends:2013}. Each Q-TLS is characterized by a~\mbox{$3$-tuple} of fundamental parameters, $\left( f_{\text{Q-TLS}} , g , \Gamma_1^{\text{Q-TLS}} \right)$. We pick~$f_{\text{Q-TLS}}$ uniformly at random from a frequency range relevant to our experiments. Since~$f_{\text{q}} \sim \SI{4.5}{\giga\hertz}$, we generate Q-TLSs with~$f_{\text{Q-TLS}} \in [ 4 , 5 ] \, \si{\giga\hertz}$.

In order to generate~$g$, we need a numerical value for both the effective electric dipole moment~$\tilde{p}$~\footnote{The angle~$\eta$ between~$\vec{p}$ and $\vec{E}_{\text{q}}$ is integrated in the distribution for~$\tilde{p}$, i.e., $\tilde{p} = \norm*{\vec{p}} \cos \eta$~\cite{Martinis:2005}.} and $\norm*{\vec{E}_{\text{q}}}$ at the position of the Q-TLS.

We pick~$\tilde{p}$ from a known probability density that has been experimentally measured, e.g., in the work of Ref.~\cite{Martinis:2005},
\begin{align}
	f_{\tilde{p}} =
		\begin{cases}
			\dfrac{1}{\tilde{p}} \sqrt{1 - \left( \dfrac{\tilde{p}}{\tilde{p}_{\text{max}}} \right)^2} ,
			& \text{for} \enskip \tilde{p}_{\text{min}} \leq \tilde{p} \leq \tilde{p}_{\text{max}} ;
\\[5.0mm]
			0 ,
			& \text{otherwise} .
		\end{cases}
	\label{Eq:ftildep}
	\raisetag{2.5mm}
\end{align}
In this equation, we set the minimum and maximum value of~$\tilde{p}$ to be~$\tilde{p}_{\text{min}} = \SI{0.1}{debye}$ and $\tilde{p}_{\text{max}} = \SI{6}{debye}$; we choose~$\tilde{p}_{\text{max}}$ as in Ref.~\cite{Barends:2013} and $\tilde{p}_{\text{min}}$ assuming that any smaller dipole moment is negligible.

The position of a Q-TLS can be randomly picked at any point within the qubit oxide layers. We may then determine~$\vec{E}_{\text{q}}$ at each of these points by means of a conformal mapping technique. This technique allows us to transform the electric field of the qubit capacitor, $\vec{E}_{\text{q}}$, into the known field of a parallel-plate capacitor. Details on this procedure are given in App.~\ref{App:QUBIT:ELECTRIC:FIELD}.

Finally, we assume that~$\Gamma_1^{\text{Q-TLS}} \propto \Delta^2_0$~\cite{Rosen:2019}, where the tunneling energy of the Q-TLS, $\Delta_0$, is picked from an inverse probability distribution. We choose the bounds such that the resulting decay rates range between~$1$ and \SI{100}{\mega\hertz}, with most rates at the low end of this range.

In order to complete step (1), we need to know the total number of Q-TLSs, $N_{\text{Q-TLS}}$, and their associated~\mbox{$3$-tuple} parameters. The Q-TLSs are hosted within an interaction region with volume determined by the length of the two CPW segments forming the qubit Al island (see App.~\ref{App:DEVICE:AND:SETUP}) and the same cross-sectional area used to pick~$\vec{E}_{\text{q}}$ (see App.~\ref{App:QUBIT:ELECTRIC:FIELD}), $V_{\text{int}} = \SI{96}{\micro\meter} \times \SI{3}{\nano\meter} \times \SI{376}{\micro\meter} \times 2$. Given a Q-TLS bandwidth~$B_{\text{Q-TLS}} = \SI{1}{\giga\hertz}$, assuming a Q-TLS density~$D = \SI{200}{\per\giga\hertz\per\micro\meter\cubed}$ (see Subsec.~\ref{Subsec:DISCUSSION:Density:of:TLSs}), and disregarding all Q-TLSs with~$g < \SI{70}{\kilo\hertz}$, we obtain~$N_{\text{Q-TLS}} \sim 570$.

In step (2), each T-TLS is characterized by a~\mbox{$2$-tuple} of fundamental parameters, $\left( \delta \! f^{\mp} , \gamma \right)$. We generate~$\delta \! f^{\mp}$ from Eq.~(\ref{Eq:hdeltafmp}), where~$\Delta$ and $\Delta_0$ are picked from the GTM distribution of Eq.~(\ref{Eq:fDeltaDelta0:GTM}). We assume~$\mathcal{E}_{\text{min}} = \SI{125}{\mega\hertz}$, $\mathcal{E}_{\text{max}} = \SI{1}{\giga\hertz}$, and $\mu = 0.3$~\cite{Faoro:2015}. The interaction energy~$U(r)$ is calculated from Eq.~(\ref{Eq:Ur}), where~$U_0 = k_{\text{B}} \times \SI{10}{\kelvin\nano\meter\cubed}$ and $r$ is the Q-TLS--T-TLS distance; this distance must be picked at random. Given a cylindrical region with radius~$r$ and height~$t_{\text{ox}}$ centered on the Q-TLS and a uniform T-TLS density, the CDF for the number of T-TLSs is proportional to~$r^2$. As a consequence, the PDF is linear in~$r$, $f_r \propto r$. We pick~$r$ from~$f_r$ assuming~$r_{\text{min}} = \SI{15}{\nano\meter}$ and $r_{\text{max}} = \SI{60}{\nano\meter}$ as bounds (see Subsec.~\ref{Subsec:DISCUSSION:Density:of:TLSs} for a discussion on~$r_{\text{max}}$).

We then generate~$\gamma$ from Eq.~(\ref{Eq:gamma}). In addition to the parameters used to generate~$\delta \! f^{\mp}$, we need~$T = \SI{60}{\milli\kelvin}$, $\gamma_0 \approx \SI{0.4}{\hertz}$, $\Omega_0 = \SI{1}{\giga\hertz}$, $m = \SI{16}{\atomicmassunit}$, and $d = \SI{2}{\angstrom}$ (see Subsec.~\ref{Subsec:DISCUSSION:Physical:Characteristics:of:a:T-TLS} for a discussion on the physical meaning of these parameters). Note that the effective qubit temperature~$T = \SI{60}{\milli\kelvin}$ corresponds to a qubit ground state population of~\SI{2.7}{\percent}, which is approximately the value observed in our experiments.

Similarly to step (1), in order to complete step (2) we need to select the number of T-TLSs interacting with each Q-TLS, $N_{\text{T-TLS}}$. We generate a set of~$N_{\text{T-TLS}} = 10$ T-TLSs, ensuring that each of them additionally fulfills the condition~$E^{+}_0 - E^{-}_0 = \sqrt{E^2_{\text{T-TLS}} + 4 U (\Delta + U)} < E_{\text{max}} = k_{\text{B}} T / 2$. We choose half of the thermal energy as our activation threshold, although similar values would work as well.

In step (3), we generate the simulated spectrotemporal charts for~$\Gamma_1^{\text{q}}$ (and, thus, $T_1$). Stochastic fluctuations are due to a T-TLS switching state randomly between the left and right well. We simulate these fluctuations as an RTS with a single~$\gamma$ for both the left and right well, i.e., assuming a symmetric noise process. For an RTS, the probability of spending a time~$t$ in a certain state is given by the PDF~$f_t = \gamma \exp(-\gamma t)$. Starting from a random state, we produce a list of times spent in each T-TLS state until reaching~$t_{\text{obs}}$. In order to generate a time series for the T-TLS state, we sample the time list at~$\Delta t$ intervals. The values of both~$\Delta t$ and $t_{\text{obs}}$ used in the simulations are the same as for the experiments and are reported in Table~\ref{Tab:Experimental:Parameters}.

The T-TLS state corresponds to a particular~$\delta f^{\mp}$. Therefore, as explained in Subsec.~\ref{Subsec:Qubit:Stochastic:Fluctuations}, the time series~$f_{\text{Q-TLS}}(t)$ for each Q-TLS can be calculated by means of Eq.~(\ref{Eq:fQ-TLSt}). Finally, we evaluate Eq.~(\ref{Eq:Gamma1q}) for all values of interest of~$f_{\text{q}}$; in order to match the spectrotemporal charts measured in the experiment, we choose~$f_{\text{q}}$ for the ranges and $N_f$ values reported in Table~\ref{Tab:Experimental:Parameters}.

The simulations are performed using the Julia Programming Language~\cite{Julia:2017}. The computer code QubitFluctuations.jl can be obtained from a GitLab repository~\cite{QubitFluctuations}.

\section{RESULTS}
	\label{Sec:RESULTS}

\begin{figure*}[ht!]
	\centering
	\includegraphics{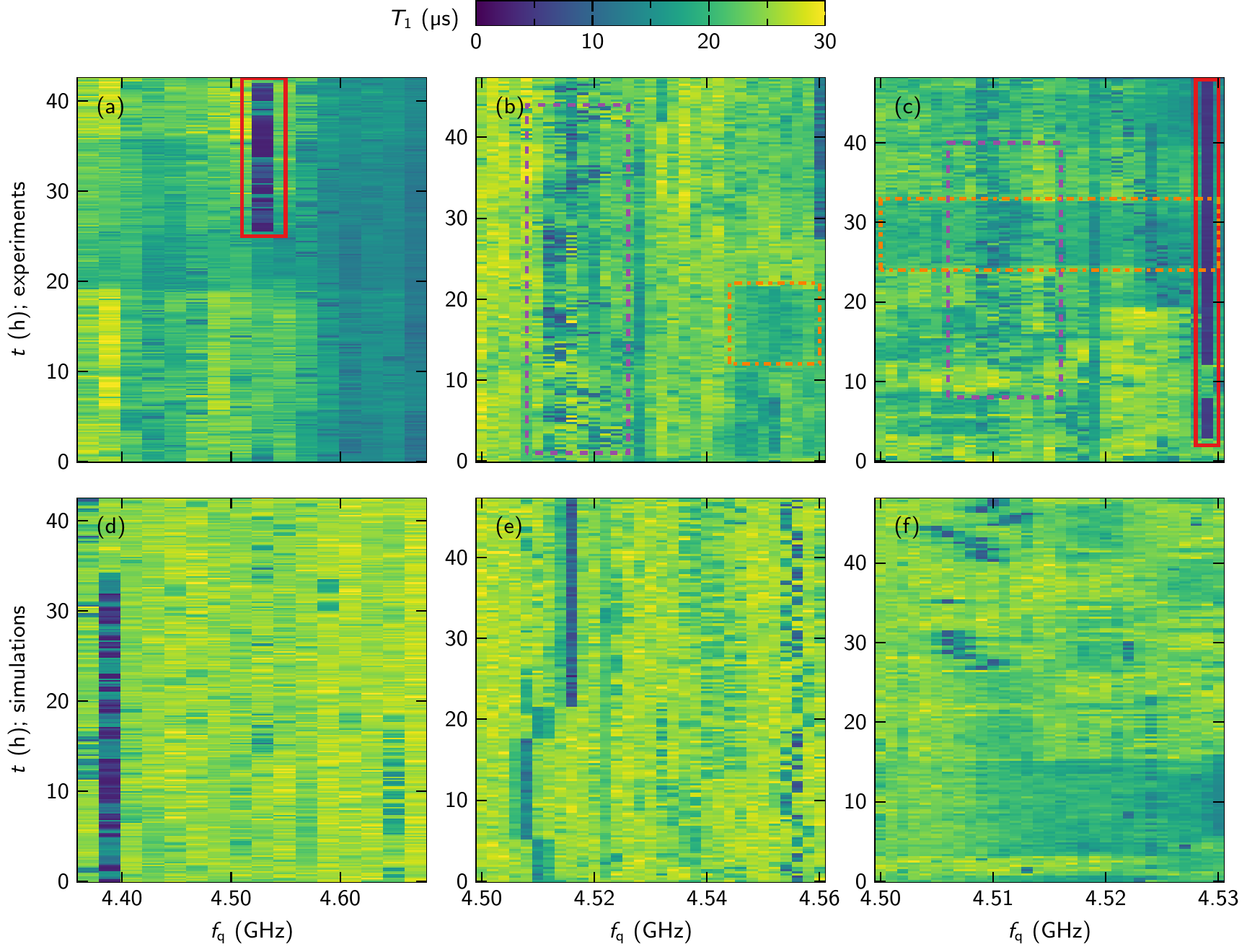}
		\caption{Experimental [(a), (b), and (c)] and simulated [(d), (e), and (f)] spectrotemporal charts of~$T_1$ vs.~$f_{\text{q}}$ and $t$, where the panels in each column display an experiment and the corresponding simulation. Spectral-diffusion patterns in the experiments are highlighted with boxes. Band-limited diffusive: dashed purple boxes. Fast narrowband telegraphic: solid red boxes. Slow wideband telegraphic: dash-dotted orange boxes. In the simulations, we add a background time series of Gaussian white noise with a standard deviation of~\SI{2}{\kilo\hertz}, which is comparable to the fitting error of our~$T_1$ experiments. \label{Fig:RESULTS:Spectrotemporal:Charts}}
\end{figure*}

The main results of this work are presented in Fig.~\ref{Fig:RESULTS:Spectrotemporal:Charts}, which shows the experimental and simulated spectrotemporal charts of~$T_1$. Details on the experiments and simulations are described in Subsecs.~\ref{Subsec:METHODS:Experiments} and \ref{Subsec:METHODS:Simulations}, respectively, with parameters reported in Table~\ref{Tab:Experimental:Parameters}. Each realization of a simulation is random due to the very nature of the method (because, e.g., $f_{\text{Q-TLS}}$ is distributed uniformly). We thus choose to display simulated spectrotemporal charts that resemble the experiments.

A visual inspection of the~$T_1$ stochastic fluctuations in Fig.~\ref{Fig:RESULTS:Spectrotemporal:Charts} reveals three distinct spectral-diffusion patterns:
\begin{enumerate}[(1)]
\item Band-limited diffusive.
\item Fast narrowband telegraphic.
\item Slow wideband telegraphic.
\end{enumerate}
Generally, it is also possible to observe combinations of such patterns.

The three patterns can be qualitatively explained by performing ad hoc simulations using a similar method as in Subsec.~\ref{Subsec:METHODS:Simulations}. However, instead of randomly generating the~$3$- and \mbox{$2$-tuple} of steps (1) and (2), we set these tuples by hand. We simulate the effect of several T-TLSs on one Q-TLS, considering three T-TLS sets with different ranges of~$\delta \! f^{\mp}$ and $\gamma$. For clarity, we choose three Q-TLSs with distinct values of~$f_{\text{Q-TLS}}$ --- Q-TLS~1, 2, and 3 --- one for each set of T-TLSs.

In broad strokes, the band-limited diffusive process is reproduced by simulating the effect of many ($\sim 10$) T-TLSs on Q-TLS~1; we select T-TLSs with high values of~$\gamma$ (ranging between tens of minutes and hours) and small values of~$\delta \! f^{\mp}$ ($< \SI{1}{\mega\hertz}$). The fast narrowband telegraphic process, instead, is generated by considering a few ($\lesssim 3$) T-TLSs acting on Q-TLS~2; in this case, we select low values of~$\gamma$ (on the order of hours) as well as small values of~$\delta \! f^{\mp}$ ($< \SI{1}{\mega\hertz}$). Similarly to the case of the fast narrowband process, the slow wideband telegraphic process is created assuming also a few ($\lesssim 3$) T-TLSs, this time coupled to Q-TLS~3; in this instance, however, we select very low values of~$\gamma$ (on the order of days) and large values of~$\delta \! f^{\mp}$ ($\leq \SI{20}{\mega\hertz}$).

\begin{table}[b!]
	\caption{T-TLS and Q-TLS parameters used in the simulations of Fig.~\ref{Fig:RESULTS:Spectral:Diffusion:Patterns}. \label{Tab:T-TLS--Q-TLS:Parameters}}
\begin{center}
	\begin{ruledtabular}
		\begin{tabular}{lcccc}
			\raisebox{0mm}[3mm][0mm]{Q-TLS} & $f_\text{Q-TLS}$ & $\Gamma_1^{\text{Q-TLS}}$ & $\gamma$ & $\delta \! f^{\mp}$ \\
			& \footnotesize{(\si{\giga\hertz})} & \footnotesize{(\si{\mega\hertz})} & \footnotesize{(\si{\hertz})} & \footnotesize{(\si{\mega\hertz})}
\\[0.5mm]
\hline\\[-2.5mm]
			\multirow{8}{*}{$1$} & \multirow{8}{*}{$4.510$} & \multirow{8}{*}{$10$} & $2 \times 10^{-5}$ & $0.9$ \\
			& {} & {} & $5 \times 10^{-5}$ & $0.7$ \\
			& {} & {} & $8 \times 10^{-5}$ & $0.7$ \\
			& {} & {} & $1 \times 10^{-4}$ & $0.6$ \\
			& {} & {} & $2 \times 10^{-4}$ & $0.6$ \\
			& {} & {} & $3 \times 10^{-4}$ & $0.5$ \\
			& {} & {} & $4 \times 10^{-4}$ & $0.3$ \\
			& {} & {} & $1 \times 10^{-3}$ & $0.1$ \\
\hline\\[-2.5mm]
			\multirow{3}{*}{$2$} & \multirow{3}{*}{$4.531$} &  \multirow{3}{*}{$5$} & $3 \times 10^{-5}$ & $0.8$ \\
			& {} & {} & $8 \times 10^{-5}$ & $0.2$ \\
			& {} & {} & $2 \times 10^{-4}$ & $0.1$ \\
\hline\\[-2.5mm]
			\multirow{2}{*}{$3$} & \multirow{2}{*}{$4.570$} & \multirow{2}{*}{$90$} & $6 \times 10^{-6}$ & $20$ \\
			& {} & {} & $8 \times 10^{-6}$ & $3$ \\[-0.5mm]
		\end{tabular}
	\end{ruledtabular}
\end{center}
\end{table}

\begin{figure*}[ht!]
	\centering
	\includegraphics{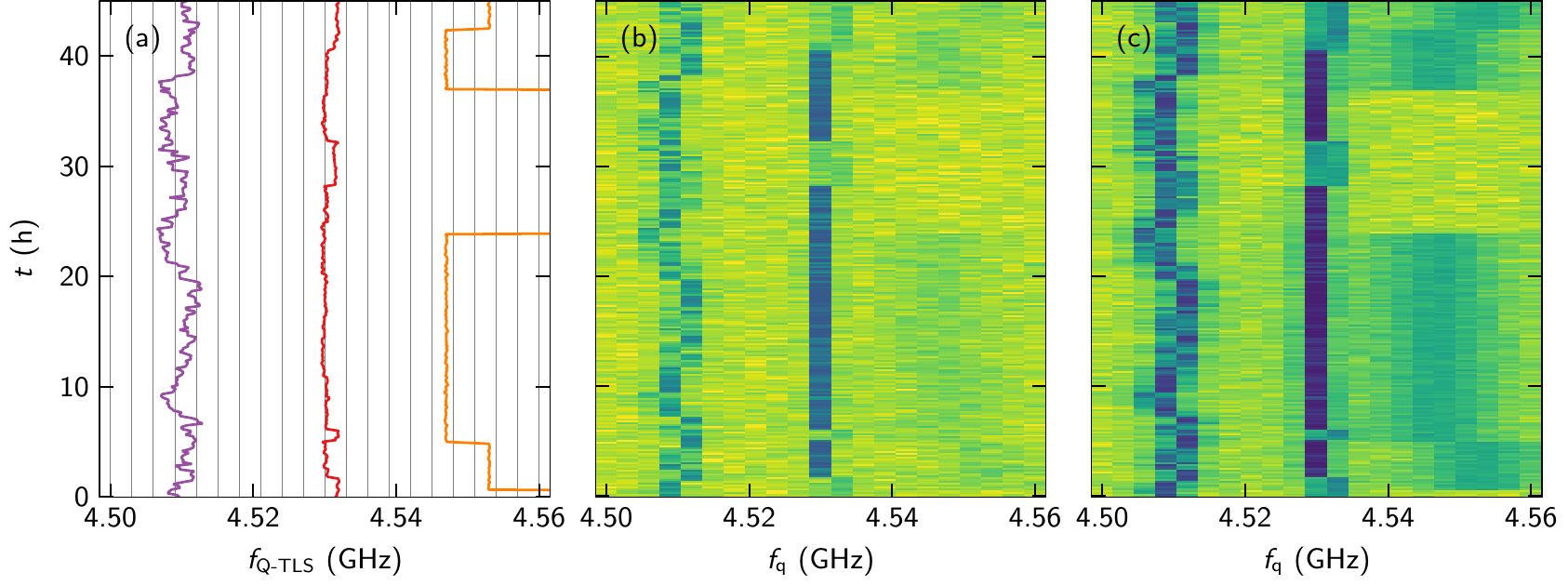}
	\caption{Three spectral-diffusion patterns. (a) Q-TLS frequency~$f_{\text{Q-TLS}}$ vs.~$t$ for Q-TLS~1 (left purple line), 2 (middle red line), and 3 (right orange line). (b) and (c) Simulated spectrotemporal charts of~$T_1$ vs.~$f_{\text{q}}$ and $t$ for~$g = 50$ and \SI{100}{\kilo\hertz}, respectively. The color map for~$T_1$ is the same as in Fig.~\ref{Fig:RESULTS:Spectrotemporal:Charts}. \label{Fig:RESULTS:Spectral:Diffusion:Patterns}}
\end{figure*}

Figure~\ref{Fig:RESULTS:Spectral:Diffusion:Patterns} illustrates the results of the simulation of the three patterns. Figure~\ref{Fig:RESULTS:Spectral:Diffusion:Patterns}~(a) exemplifies the effect of the three different sets of T-TLSs on Q-TLS~1, 2, and 3. Figures~\ref{Fig:RESULTS:Spectral:Diffusion:Patterns}~(b) and (c) demonstrate the impact of each Q-TLS on the spectrotemporal chart of~$T_1$ for a small~(a) and large~(b) value of~$g$. The T-TLS and Q-TLS parameters used in the simulations are reported in Table~\ref{Tab:T-TLS--Q-TLS:Parameters}.

Q-TLS~1 is affected by many T-TLSs that switch continuously within the observation time. The T-TLSs act additively on the Q-TLS, resulting in a diffusive shift of~$f_{\text{Q-TLS}}$ [see Eq.~(\ref{Eq:fQ-TLSt})]. Different from Brownian diffusion, the shift in~$f_{\text{Q-TLS}}$ does not exceed the sum of the individual frequency shifts induced by each T-TLS at any observation time. The diffusive process is thus characterized by a limited frequency bandwidth, as shown in Fig.~\ref{Fig:RESULTS:Spectral:Diffusion:Patterns}~(a). The spectrotemporal chart of~$T_1$ displays a similar behavior; $T_1$ fluctuates in time over a finite frequency range, exhibiting moderate and strong variations in Figs.~\ref{Fig:RESULTS:Spectral:Diffusion:Patterns}~(b) and (c), respectively.

Q-TLS~2, which is affected by a few T-TLSs, switches mainly between two values of~$f_{\text{Q-TLS}}$ (low and high); for both states, much smaller fluctuations at higher switching rates are noticeable. The telegraphic nature of this process affects dramatically the spectrotemporal chart of~$T_1$ when~$f_{\text{Q-TLS}} \simeq f_{\text{q}}$. This is the case in the example of Fig.~\ref{Fig:RESULTS:Spectral:Diffusion:Patterns}~(a) when Q-TLS~2 dwells in the low frequency position. In this state, $T_1$ becomes largely reduced compared to when the Q-TLS resides in the high frequency position, as displayed in Figs.~\ref{Fig:RESULTS:Spectral:Diffusion:Patterns}~(b) and (c). The low value of~$\Gamma_1^{\text{Q-TLS}}$ leads to a narrowband process, with more pronounced~$T_1$ variations in Fig.~\ref{Fig:RESULTS:Spectral:Diffusion:Patterns}~(b) compared to Fig.~\ref{Fig:RESULTS:Spectral:Diffusion:Patterns}~(c).

It is worth noting that, in our example, the high frequency position lies between two values of~$f_{\text{q}}$ [vertical solid light-gray lines in Fig.~\ref{Fig:RESULTS:Spectral:Diffusion:Patterns}~(a)] but is too far from either of them to significantly impact~$T_1$. This effect shows that the frequency resolution of our experiments [i.e., the~$x$-axis ``pixeling'' in Figs.~\ref{Fig:RESULTS:Spectral:Diffusion:Patterns}~(b) and (c)] affects the spectrotemporal chart of~$T_1$.

Q-TLS~3 behaves similarly to Q-TLS~2, although one of the T-TLSs has a significantly larger value of~$\delta \! f^{\mp}$. Due to low values of~$\gamma$, Q-TLS~3 undergoes telegraphic frequency shifts only a couple of times during observation. The high value of~$\Gamma_1^{\text{Q-TLS}}$ strongly damps the effect on~$T_1$, resulting in a wideband process. In fact, the effect is barely visible in Fig.~\ref{Fig:RESULTS:Spectral:Diffusion:Patterns}~(b), even when the~Q-TLS is almost on resonance with the qubit. In presence of a strong coupling, however, the impact on the spectrotemporal chart of~$T_1$ is clearly identifiable; as shown in Fig.~\ref{Fig:RESULTS:Spectral:Diffusion:Patterns}~(c), the effect extends over a large frequency range.

\section{DISCUSSION}
	\label{Sec:DISCUSSION}

In this section, we discuss the physical characteristics of a T-TLS (Subsec.~\ref{Subsec:DISCUSSION:Physical:Characteristics:of:a:T-TLS}); we then discuss the density of TLSs (Subsec.~\ref{Subsec:DISCUSSION:Density:of:TLSs}); finally, we provide insight on the interpretation of the Allan deviation and power spectral density (Subsec.~\ref{Subsec:DISCUSSION:On:the:Interpretation:of:the:Allan:Deviation:and:Power:Spectral:Density}).

\subsection{Physical Characteristics of a T-TLS}
	\label{Subsec:DISCUSSION:Physical:Characteristics:of:a:T-TLS}

The two quantities required to represent T-TLSs in the simulations shown in Fig.~\ref{Fig:RESULTS:Spectrotemporal:Charts} are~$\delta \! f^{\mp}$ and $\gamma$ of Eqs.~(\ref{Eq:hdeltafmp}) and (\ref{Eq:gamma}), respectively. The former is determined only by parameters chosen according to the GTM. The latter requires the knowledge of additional physical characteristics of T-TLSs: $m$ and $d$, as well as $\Omega_0$.

We assume that TLSs, and thus T-TLSs, are hosted in \emph{oxide} layers at the SM, SA, or MA interfaces (see Sec.~\ref{Sec:INTRODUCTION}). The oxide layers are composed of molecules with an oxygen~(O) atom bound to a pair of neighboring atoms. A T-TLS can be modeled as an \emph{O atom} with mass~$m = \SI{16}{\atomicmassunit}$ tunneling between two wells (i.e., states) at a distance~$d$ from each other. It is reasonable to assume that~$d$ is comparable to the bond length between the~O atom and a neighboring atom~\cite{Phillips:1987}. In many applications, using~Si or sapphire substrates and Al as a metal results in amorphous Si or Al oxide interfacial layers. The bond length between the~O and Si or Al atoms is on the order of~\SI{2}{\angstrom}~\cite{Wells:1984,Song:2016}; this is why in our simulations we choose~$d = \SI{2}{\angstrom}$.

Equation~(\ref{Eq:Delta0}) is valid only when~$V \geq 0$. Accordingly, it must be that~$\Omega_0 \geq \Delta_0$ for all values of~$\Delta_0$ picked from the GTM distribution. On the one hand, choosing a value~$\Omega_0 \sim \Delta_0$ leads to~$V \sim 0$, which would correspond to a single- rather than a double-well potential. On the other hand, we cannot choose~$\Omega_0$ to be arbitrarily large due to its relationship to~$\gamma$,
\begin{equation}
\gamma = \gamma_0 \, \exp\left[-\left(\dfrac{\hbar}{d}\right)^{\!\! 2} \dfrac{1}{2 m} \left(\ln{\dfrac{\Omega_0}{\Delta_0}}\right)^{\!\! 2} \Big/ (k_{\text{B}} T)\right] .
\end{equation}
In fact, there is a small range of values of~$\Omega_0$ that results in a distribution of~$\gamma$ similar to that empirically inferred from the spectrotemporal charts of Fig.~\ref{Fig:RESULTS:Spectrotemporal:Charts}. We choose~$\Omega_0 = \SI{1}{\giga\hertz}$ to match the experimental range~$\gamma \in [10^{-6} , 10^{-2}]~\si{\hertz}$ (i.e., from days to minutes) as closely as possible. In this case, we obtain T-TLSs with~$V \gtrsim \SI{1.8}{\giga\hertz}$.

\subsection{Density of TLSs}
	\label{Subsec:DISCUSSION:Density:of:TLSs}

The TLS density~$D$ is estimated by counting the number~$N$ of TLSs within a certain interaction region with volume~$V_{\text{int}}$ and bandwidth~$B$, $D = N / ( V_{\text{int}} \, B )$.

In the case of Q-TLSs, their number~$N_{\text{Q-TLS}}$ can be readily obtained by counting the interactions between a qubit and a Q-TLS in spectroscopy experiments~\cite{Barends:2013,Bejanin:2020,Bilmes:2021,Mamin:2021}. For qubits where Q-TLSs are hosted in a volume of native oxide, the estimated density is~$D_{\text{Q-TLS}} \sim \SI{100}{\per\giga\hertz\per\micro\meter\cubed}$. In order to reproduce well our experimental spectrotemporal charts, in the simulations we choose~$D_{\text{Q-TLS}} = \SI{200}{\per\giga\hertz\per\micro\meter\cubed}$.

Spectroscopic methods cannot be used to count the number of T-TLSs because, at such low frequencies, the qubit is in an incoherent thermal state. The experimental spectrotemporal charts reveal that Q-TLSs are generally affected by multiple sources of telegraphic noise, as clearly shown by the band-limited diffusive pattern in Fig.~\ref{Fig:RESULTS:Spectrotemporal:Charts}. This observation makes it possible to infer the number of T-TLSs coupled to each Q-TLS, $N_{\text{T-TLS}}$; in the simulations, we choose~$N_{\text{T-TLS}} = 10$. These T-TLSs are assumed to be contained inside an interaction region with volume~$V_{\text{int}}$ centered on their host Q-TLS. It is worth pointing out that our choice of~$N_{\text{T-TLS}} = 10$ can still result in both the fast narrowband and slow wideband telegraphic patterns in Fig.~\ref{Fig:RESULTS:Spectrotemporal:Charts}; this is because $\delta f^{\mp}$ and $\gamma$ are distributed over a large parameter range possibly leading to a single predominant T-TLS.

The experiment of Fig.~\ref{Fig:RESULTS:Spectrotemporal:Charts}~(c) allows us to resolve T-TLSs with interaction strengths~$U(r) \geq \SI{1}{\mega\hertz}$. According to Eq.~(\ref{Eq:Ur}), this condition corresponds to a maximum interaction distance~$r_{\text{max}} = \SI{60}{\nano\meter}$. Notably, this condition is similar to that hypothesized in the work of Ref.~\cite{Faoro:2015}. As explained in Subsec.~\ref{Subsec:METHODS:Simulations}, the T-TLS interaction region is a cylinder with radius~$r_{\text{max}}$ and a height of~$t_{\text{ox}}$; the volume associated with this region is~$V_{\text{int}} \approx \SI{3.4e-05}{\micro\meter\cubed}$~\footnote{For the Q-TLS density used in our simulations, $D_{\text{Q-TLS}} = \SI{200}{\per\giga\hertz\per\micro\meter\cubed}$, we can find a Q-TLS area density~$\sigma_{\text{Q-TLS}} = D_{\text{Q-TLS}} \times \SI{1}{\giga\hertz} \times \SI{3}{\nano\meter} = \SI{0.6}{\per\micro\meter\squared}$. The average area per Q-TLS is therefore~$1 / \sigma_{\text{Q-TLS}}$. Assuming each Q-TLS is contained within a square, the radius of the circle inscribed in each square is~$r_{\text{Q-TLS}} = \sqrt{1 / \sigma_{\text{Q-TLS}}} / 2 \approx \SI{600}{\nano\meter}$. Since~$r_{\text{max}} \ll r_{\text{Q-TLS}}$, the T-TLS interaction regions do not overlap on average and, thus, we are not double counting T-TLSs.}.

Given~$B = ( E_{\text{max}} - E_{\text{min}} ) / h = \SI{500}{\mega\hertz}$, we finally obtain~$D_{\text{T-TLS}} \approx \SI{6e+05}{\per\giga\hertz\per\micro\meter\cubed}$. This value is much larger than~$D_{\text{Q-TLS}}$, suggesting that~$D$ varies significantly in frequency and is higher at lower frequencies. This finding is in contrast with the typical assumption made by the STM practitioners that TLSs are uniformly distributed in frequency. It is worth noting that a result similar to ours has been recently reported in the work of Ref.~\cite{Klimov:2018}, although our value for~$D_{\text{T-TLS}}$ is even larger than in that work.

\subsection{On the Interpretation of the Allan Deviation and Power Spectral Density}
	\label{Subsec:DISCUSSION:On:the:Interpretation:of:the:Allan:Deviation:and:Power:Spectral:Density}

\begin{table}[b!]
	\caption{Time-series simulation parameters used in Fig.~\ref{Fig:AD:PSD}. The simulations are performed as described in Subsec.~\ref{Subsec:METHODS:Simulations}; however, instead of randomly picking all relevant parameters, we manually specify them. Note that~$\gamma = 1 / (2 \tau_0)$. \label{Tab:Q-TLS:T-TLS:Parameters}}
\begin{center}
	\begin{ruledtabular}
		\begin{tabular}{lccccc}
			$M_\text{T-TLS}$ & $f_\text{Q-TLS}$ & $g$ & $\Gamma^\text{Q-TLS}_1$ & $\gamma$ & $\delta\! f^\mp$ \\
			& \footnotesize{(\si{\giga\hertz})} & \footnotesize{(\si{\mega\hertz})} & \footnotesize{(\si{\mega\hertz})} & \footnotesize{(\si{\micro\Hz})} & \footnotesize{(\si{\mega\hertz})} \\[0.5mm]
\hline \\[-3mm]
			\multirow{1}{*}{$1$} & $4.5011$ & $0.04$ & $15$ & $100$ & $0.6$ \\
\hline \\[-3mm]
			\multirow{4}{*}{$4$} & $4.5011$ & $0.02$ & $10$ & $75$ & $0.8$ \\
			& $4.5015$ & $0.02$ & $10$ & $70$ & $0.6$ \\
			& $4.4989$ & $0.02$ & $10$ & $140$ & $0.8$ \\
			& $4.4986$ & $0.02$ & $10$ & $75$ & $0.4$ \\[-0.5mm]
		\end{tabular}
	\end{ruledtabular}
\end{center}
\end{table}

\begin{figure}[t!]
	\centering
	\includegraphics{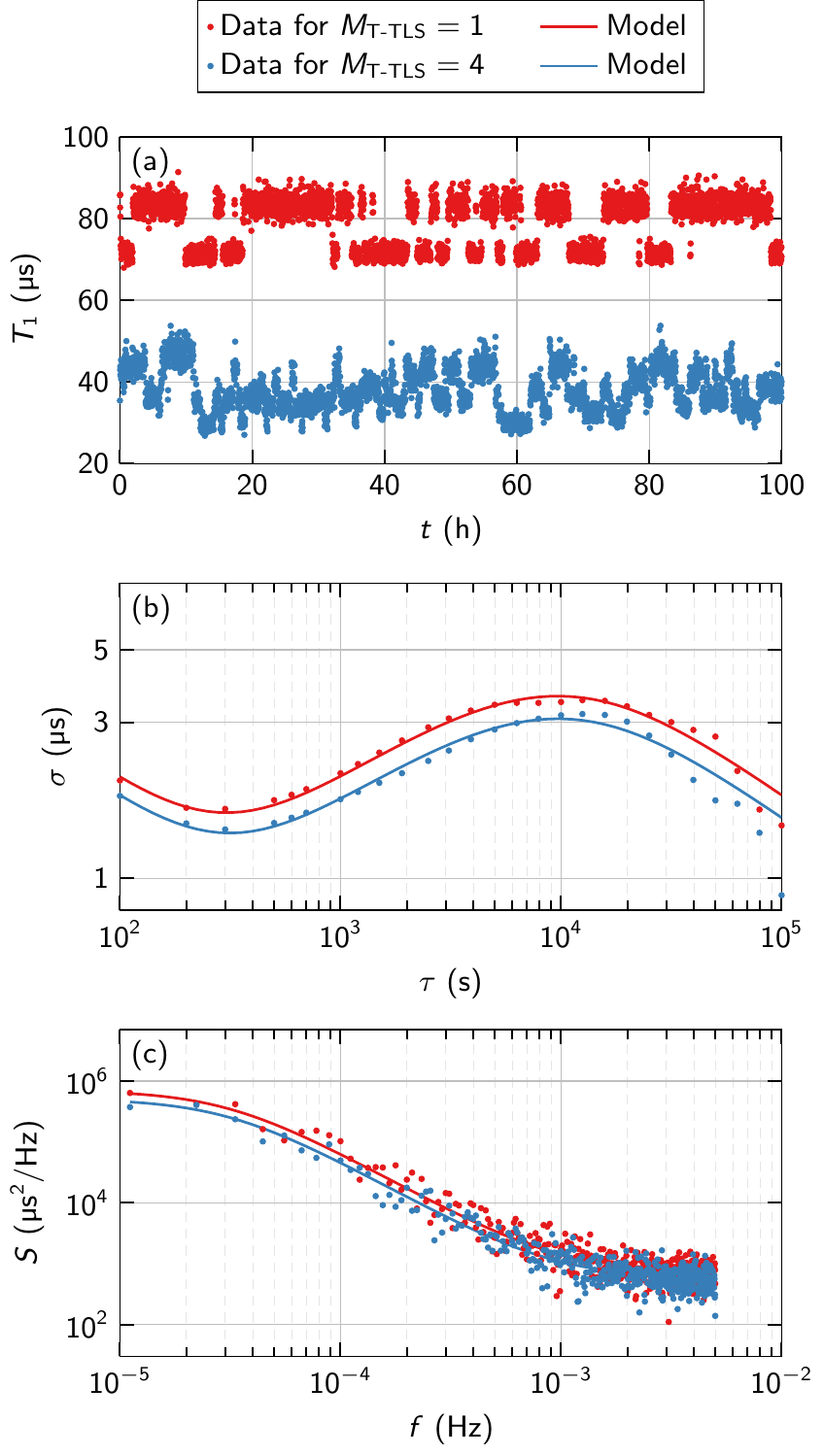}
	\caption{Comparison between the statistical analyses of two simulated times series. (a) Simulated time series of~$T_1$ vs.~$t$. The series for~$M_{\text{T-TLS}} = 1$ is vertically offset by~\SI{40}{\micro\second} for clarity. (b) Estimated AD~$\sigma$ vs.~$\tau$ and associated fitting curves from Eq.~(\ref{Eq:sigmasquared}). We find~$A_0 = 5.84(5)$ and \SI{4.98(6)}{\micro\second}, $h_0 = 749(111)$ and \SI{583(119)}{\micro\second\squared\per\hertz}, and $1 / \tau_0 = 195(7)$ and \SI{195(10)}{\micro\hertz} for~$M_{\text{T-TLS}} = 1$ and $4$, respectively. Note that we are fitting~$\sigma^2$ with the Levenberg–Marquardt algorithm, but plotting~$\sigma$. The AD is computed at logarithmically-spaced points. (c) Estimated PSD~$S$ vs.~$f$. We use the fitting parameters from~(b) to overlay the model of Eq.~(\ref{Eq:S}) to the data. The PSD is estimated using the Welch's method with~\SI{25}{\hour} overlapping segments (rectangular window). The value of~$\tau_0$ fitted for~$M_{\text{T-TLS}} = 1$ matches (within the confidence interval) that chosen in the simulations and reported in Table~\ref{Tab:Q-TLS:T-TLS:Parameters}; the fitted~$\tau_0$ for~$M_{\text{T-TLS}} = 4$, instead, does not match any of the values in Table~\ref{Tab:Q-TLS:T-TLS:Parameters}. \label{Fig:AD:PSD}}
\end{figure}

Time series experiments similar to those reported here are frequently studied by means of statistical analyzes such as the Allan deviation~(AD) or the power spectral density~(PSD), or both. For example, this approach has been pursued in the work of Refs.~\cite{Burnett:2019,Schloer:2019}. It is tempting to ascribe simple models to these statistical estimators in order to extract T-TLS parameters such as their switching rate~$\gamma$ and number~$M_{\text{T-TLS}}$; in this case, $M_{\text{T-TLS}}$ is the total number of T-TLSs affecting the qubit by interacting with a single or multiple Q-TLSs. It is common, however, to encounter scenarios where these models are misleading.

Figure~\ref{Fig:AD:PSD} presents two distinct scenarios that illustrate this issue. The time series in Fig.~\ref{Fig:AD:PSD}~(a) are obtained by simulating one scenario with~$M_{\text{T-TLS}} = 1$ and another with~$M_{\text{T-TLS}} = 4$. The simulations parameters are reported in Table~\ref{Tab:Q-TLS:T-TLS:Parameters}. As expected, there is a stark visual difference between the two time series: In the first scenario, it is possible to clearly identify one RTS; this is impossible in the second scenario. However, this difference is not reflected in either the AD or PSD. In both simulated scenarios, we observe a pronounced peak in the AD and a lobe in the PSD. These features are indicative of Lorentzian noise. However, they appear to be practically the same for the two scenarios. In fact, it is possible to fit the AD or PSD using a simple model based on a single source of Lorentzian noise, along with white noise. The model reads
\begin{equation}
\sigma^2 = \dfrac{h_0}{2 \tau} + \left( \dfrac{A_0 \tau_0}{\tau} \right)^2 \left( 4 e^{-\tau / \tau_0} - e^{- 2 \tau / \tau_0} - 3 + \dfrac{2 \tau}{\tau_0} \right)
	\label{Eq:sigmasquared}
\end{equation}
for the AD and
\begin{equation}
S = h_0 + \dfrac{4 A^2_0 \tau_0}{1 + (2 \pi f \tau_0)^2}
	\label{Eq:S}
\end{equation}
for the PSD, where~$\tau$ and $f$ are the analysis interval and frequency, $h_0$ and $A_0$ are the white and Lorentzian noise amplitudes, and $\tau_0$ is the Lorentzian characteristic time~\cite{Riley:2008}.

Although the two simulated time series are associated with entirely different scenarios, the simple models of Eqs.~(\ref{Eq:sigmasquared}) and (\ref{Eq:S}) fit accurately both the AD and PSD for very similar values of~$\tau_0$; we obtain~$1 / \tau_0 = \SI{195(7)}{\micro\hertz}$ when~$M_{\text{T-TLS}} = 1$ and $1 / \tau_0 = \SI{195(10)}{\micro\hertz}$ when~$M_{\text{T-TLS}} = 4$. This conclusion can be qualitatively understood by noticing that multiple physical sources of Lorentzian noise combine to form a single wideband peak in the AD (or lobe in the PSD). As a consequence, this feature can be mistakenly fitted with a model comprising a single Lorentzian term. For this reason, we elect \emph{not to analyze} our experimental results by ascribing simple models to the AD (or PSD).

\section{CONCLUSIONS}
	\label{Sec:CONCLUSIONS}

We study the physics of TLSs by means of a frequency-tunable planar superconducting qubit. We show that simulations based on the TLS interacting model (or GTM) can explain the spectrotemporal charts of~$T_1$ observed in the experiments over long time periods. We find that the density of T-TLSs is much larger than that of Q-TLSs, meaning TLSs are nonuniformly distributed over large frequency bandwidths. Our finding corroborates the results reported in the work of Ref.~\cite{Klimov:2018}.

Our experiments demonstrate that the additional dimension provided by frequency tunability makes tunable qubits a better probe to study spectral diffusion compared to fixed-frequency devices. Hence, we suggest that future work on TLS stochastic fluctuations should explore even wider frequency bandwidths. A large bandwidth would increase the chances to encounter a scenario where a pair of Q-TLSs interacts with a single T-TLS, resulting in a synchronous fluctuation of the two Q-TLSs. Such an experiment would conclusively prove the validity of the TLS--TLS interaction hypothesis in the GTM.

Lastly, we expect that performing experiments at different operating temperatures would provide one more knob to modify the frequency bandwidth of thermally activated TLSs. This approach would allow us to explore the TLS density for different frequency ranges.

\begin{acknowledgments}
This research was undertaken thanks in part to funding from the Canada First Research Excellence Fund~(CFREF) and the Discovery and Research Tools and Instruments Grant Programs of the Natural Sciences and Engineering Research Council of Canada~(NSERC). We would like to acknowledge the Canadian Microelectronics Corporation~(CMC) Microsystems for the provision of products and services that facilitated this research, including CAD software. The authors thank the Quantum-Nano Fabrication and Characterization Facility at the University of Waterloo.
\end{acknowledgments}

\appendix

\section{DERIVATION OF~\texorpdfstring{$\Gamma_1^{\text{q,Q-TLS}}$}{GAMMA1}}
	\label{App:DERIVATION:OF:Gamma1qQTLS}

In this appendix, we derive Eq.~(\ref{Eq:Gamma1qQ-TLS}). The master equation in Lindblad form of a qubit--Q-TLS system reads
\begin{equation}
\dfrac{d \hat{\rho}}{dt} = - \dfrac{i}{\hbar} [\widehat{H}_{\text{q,Q-TLS}} , \hat{\rho}] + \sum_{j} \left( \hat{L}_j \hat{\rho} \hat{L}^{\dagger}_j - \dfrac{1}{2} \left\{\hat{L}^{\dagger}_j \hat{L}^{}_j , \hat{\rho}\right\} \right) ,
	\label{Eq:App:rhodot}
\end{equation}
where~$\hat{\rho}(t)$ is the density matrix, $\widehat{H}_{\text{q,Q-TLS}}$ is given by Eq.~(\ref{Eq:H_qQ-TLS}), $j \in \{ \text{q} , \text{Q-TLS} \}$, and $\hat{L}_j$ and $\hat{L}^{\dagger}_j$ are Lindblad operators.

Our study is focused on the fluctuations in~$T_1$. Hence, in Eq.~(\ref{Eq:App:rhodot}) we only account for the energy relaxation rates of the qubit and Q-TLS. In this case, the Lindblad operators are~$\hat{L}_{\text{q}} = \sqrt{\widetilde{\Gamma}_1^{\text{q}}} \, \hat{\sigma}_{\text{q}}^{-}$ and $\hat{L}_{\text{Q-TLS}} = \sqrt{\Gamma_1^{\text{Q-TLS}}} \, \hat{\sigma}_{\text{Q-TLS}}^{-}$.

The quantity~$\hat{L}_j \hat{\rho} \hat{L}^{\dagger}_j = 0$ at all times because there is at most one excitation in a qubit--Q-TLS coupled system. With this assumption, and by defining the effective non-Hermitian Hamiltonian~\cite{Meystre:2007}
\begin{equation}
\widehat{H}_{\text{eff}} = \widehat{H}_{\text{q,Q-TLS}} - \dfrac{i}{2} \left( \widetilde{\Gamma}_1^{\text{q}} \, \hat{\sigma}_{\text{q}}^{+} \hat{\sigma}_{\text{q}}^{-} + \Gamma_1^{\text{Q-TLS}} \, \hat{\sigma}_{\text{Q-TLS}}^{+} \hat{\sigma}_{\text{Q-TLS}}^{-} \right) ,
	\label{Eq:App:Heff}
\end{equation}
the Lindbladian of Eq.~(\ref{Eq:App:rhodot}) can be written as a simple Schr\"odinger equation with a ``decaying wave function'' $\ket{\Psi(t)} = \alpha(t) \ket{\text{e}} + \beta(t) \ket{1}$, where~$\alpha(t)$ and $\beta(t)$ are the time-dependent complex amplitudes associated with the excited state~$\ket{\text{e}}$ of the qubit and $\ket{1}$ of the Q-TLS.

The exact result of the Schr\"{o}dinger equation for~$\alpha(t)$ given that~$\alpha(t=0) = 1$ and $\beta(t=0) = 0$ is
\begin{align}
\alpha(t) = &
\dfrac{1}{2 \Lambda}
\left[
a \exp \left( - \dfrac{\Lambda}{4} t \right) +
b \exp \left( \dfrac{\Lambda}{4} t \right)
\right] \nonumber\\
& \times \exp \left( - \dfrac{\widetilde{\Gamma}_1^{\text{q}} + \Gamma_1^{\text{Q-TLS}}}{4} t \right) ,
	\label{Eq:App:alphat}
\end{align}
where~$\Lambda$ is given by Eq.~(\ref{Eq:Lambda}), $a = \Lambda - (\Gamma_1^{\text{Q-TLS}} - \widetilde{\Gamma}_1^{\text{q}}) + 4 \pi i \Delta f$, and $b = \Lambda + (\Gamma_1^{\text{Q-TLS}} - \widetilde{\Gamma}_1^{\text{q}}) - 4 \pi i \Delta f$.

Since we are calculating a decay, we are only interested in the envelope of~$\alpha(t)$, $\tilde{\alpha}(t)$. We thus set~$\Im[\Lambda] = 0$ in the two exponential terms of Eq.~(\ref{Eq:App:alphat}) and calculate the envelope probability~$\tilde{P}_{\text{e}}(t) = \abs{\tilde{\alpha}(t)}^2$ for the qubit to be in~$\ket{\text{e}}$,
\begin{align}
\tilde{P}_{\text{e}}(t) & = \abs{\dfrac{{a}}{2 \Lambda}}^2 \exp \left[ - \dfrac{\Gamma_1^{\text{Q-TLS}} + \widetilde{\Gamma}_1^{\text{q}} + \Re[\Lambda]}{2} t \right] \nonumber\\
			 			& + \abs{\dfrac{{b}}{2 \Lambda}}^2 \exp \left[ - \dfrac{\Gamma_1^{\text{Q-TLS}} + \widetilde{\Gamma}_1^{\text{q}} - \Re[\Lambda]}{2} t \right] \nonumber\\
			 			& + \dfrac{a b^{\ast} + a^{\ast} b}{\abs{2 \Lambda}^2} \exp \left[ - \dfrac{\Gamma_1^{\text{Q-TLS}} + \widetilde{\Gamma}_1^{\text{q}}}{2} t \right] .
	\label{Eq:App:tildePet}
\end{align}
When~$\Gamma_1^{\text{Q-TLS}} > \widetilde{\Gamma}_1^{\text{q}}$, which is the regime of interest in our experiments, the term proportional to~$\abs{b}^2$ in Eq.~(\ref{Eq:App:tildePet}) is dominant. Therefore, in order to find an approximate expression for the Q-TLS contribution only, we subtract the qubit contribution~$\widetilde{\Gamma}_1^{\text{q}}$ from the rate in the exponential proportional to~$\abs{b}^2$. This procedure results in Eq.~(\ref{Eq:Gamma1qQ-TLS}) in the main text.

\section{DEVICE AND SETUP}
	\label{App:DEVICE:AND:SETUP}

The superconducting Xmon transmon qubit~\cite{Barends:2013} used in this article is the same as in our work of Ref.~\cite{Bejanin:2020}, with micrographs shown in that manuscript. The qubit consists of an Al~island in parallel with a superconducting quantum interference device~(SQUID).

The Al island forms a capacitor that is composed of two intersecting CPW segments in the shape of a Greek cross, where each segment has length~$L = \SI{376}{\micro\meter}$. One segment is formed by a center conductor, or strip, of width~$S = \SI{24}{\micro\meter}$ and is separated by a distance~$W = \SI{24}{\micro\meter}$ from a ground plane on each side of the strip. The capacitance of the island is~$C_{\text{q}} \approx \SI{100}{\femto\farad}$ (corresponding to a single-electron charge energy~$E_{\text{c}} / h \approx \SI{188.6}{\mega\hertz}$).

The qubit capacitor is connected in parallel with the SQUID, which is made of an Al loop interrupted by two parallel Josephson tunnel junctions with critical current~$I_{\text{c}0} \approx \SI{17.4}{\nano\ampere}$ (corresponding to a Josephson energy~$E_{\text{J}} / h \approx \SI{8.6}{\giga\hertz}$) for each junction. The SQUID forms the inductive element of the qubit.

Due to the SQUID design, we are able to tune the SQUID critical current~$I_{\text{c}}$ in situ during the experiment by threading the SQUID loop with a flux~$\phi_Z = M_Z \, i_Z$, where~$M_Z \sim \SI{3}{\pico\henry}$ is the mutual inductance between the loop and an external circuit with current~$i_Z$. A quasi-static flux bias~$\phi^{\text{qs}}_Z$ allows us to set the qubit frequency~$f_{\text{q}} (\phi^{\text{qs}}_Z)$, i.e., the qubit bias point. The qubit parameters given above result in a zero-bias~$f_{\text{q}} (\phi^{\text{qs}}_Z=0) \approx \SI{4.8}{\giga\hertz}$.

The qubit can be controlled by means of~$X$ or $Y$ microwave pulses, which are applied through a capacitive network with coupling capacitor of capacitance~$C_{XY} \approx \SI{100}{\atto\farad}$. The qubit state is measured by means of a readout resonator with~$f_{\text{r}} \approx \SI{5}{\giga\hertz}$, which is capacitively coupled with a coupling capacitor of capacitance~$C_M \approx \SI{3.4}{\femto\farad}$. We read out the qubit state over~$655$ single-shot measurements to find~$P_{\text{e}}$ with a visibility~$\gtrsim \SI{90}{\percent}$.

The qubit is fabricated by depositing and patterning thin-film Al on thoroughly cleaned surfaces; we use the same cleaning process as in our work of Ref.~\cite{Earnest:2018}. The Josephson tunnel junctions are fabricated using a standard double-angle Niemeyer–Dolan technique. The qubit is operated at the base temperature of a dilution refrigerator, approximately~\SI{10}{\milli\kelvin}. The control and measurement signals are applied through a heavily filtered microwave network. The setup is the same as in our work of Ref.~\cite{Bejanin:2020}, which shows a detailed diagram of the control and measurement lines.

\section{EXPERIMENTAL DETAILS}
	\label{App:EXPERIMENTAL:DETAILS}

\begin{figure}[t!]
	\centering
	\includegraphics{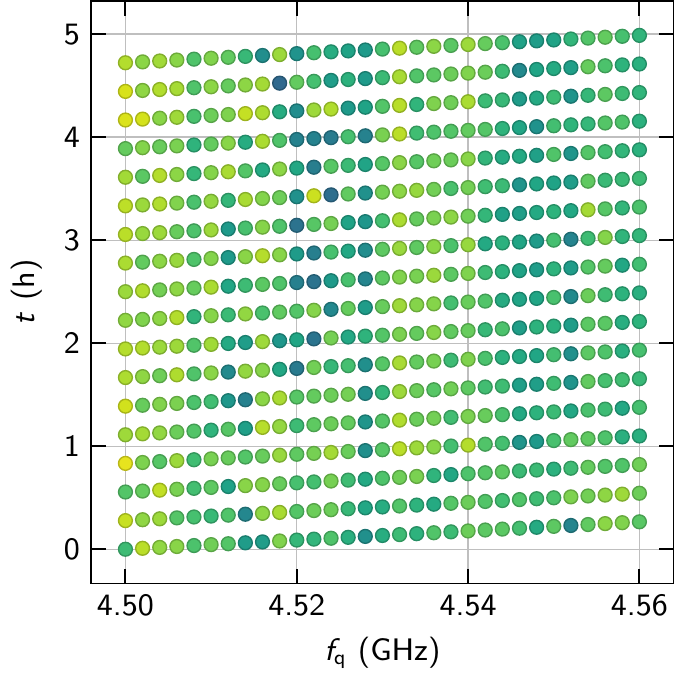}
	\caption{Scatter plot of~$T_1$ vs.~$f_{\text{q}}$ and $t$ for dataset~$2$ [same dataset as in Fig.~\ref{Fig:RESULTS:Spectrotemporal:Charts}~(c)] at the actual measurement time; the color map for~$T_1$ is the same as in Fig.~\ref{Fig:RESULTS:Spectrotemporal:Charts}. Note that the vertical axis is truncated at~$t = \SI{5}{\hour}$ to display the relative measurement times more clearly. \label{Fig:APPENDIX:Scatter:Chart}}
\end{figure}

The spectrotemporal charts displayed in Sec.~\ref{Sec:RESULTS} can be interpreted as matrices of~$T_1$ values, with~$m$~rows and $n$~columns; $m$ and $n$ represent a time and frequency index, respectively. The~$(1,1)$ entry is the bottom-left element of the matrix, such that time increases from bottom to top. We set~$f_{\text{q}}$ from low to high values, completing one row of each matrix when reaching the highest value of~$f_{\text{q}}$. Subsequent rows are measured restarting always from the lowest value of~$f_{\text{q}}$. Hence, the time~$t_{m,n}$ at which each data point~$(m,n)$ is taken increases from left to right for the~$m$-th row, starting at~$t_{m,1}$ and ending at~$t_{m,N_f}$. The time difference between subsequent rows is a constant value defined as~$\Delta t = t_{m+1,1} - t_{m,1}$. Although each measurement in any particular row is taken at a different time, we choose to display the data on a rectangular matrix where each row element is associated with the same time value. As a comparison, Fig.~\ref{Fig:APPENDIX:Scatter:Chart} shows a scatter plot for which each~$T_1$ value is plotted at the actual measurement time. This figure elucidates two limitations of our experiments: (1) The impossibility to measure an entire row at exactly the same time. (2) The fact that~$t_{m,N_f} \sim t_{m+1,1}$. It additionally stresses a difference between experiments and simulations, i.e., the fact that in simulations all row elements are calculated at the exact same time.

In order to keep~$\Delta t$ constant we must account for experimental nonidealities. The time required to perform a single~$T_1$ experiment is~$t_{\text{exp}} \approx \SI{16}{\second}$ and varies slightly between experiments. In addition, latencies in the electronic equipment when setting a new value of~$f_{\text{q}}$ result in a short time overhead. To overcome these issues, we measure a test row and record the corresponding measurement time. We then augment this measurement time by a certain buffer time, which we estimate to be sufficiently longer than any possible time variations due to nonidealities. The sum of the measurement time of the test row and the buffer time is~$\Delta t$. For example, for the dataset shown in Fig.~\ref{Fig:RESULTS:Spectrotemporal:Charts}~(c), the time elapsed to acquire the data of the test row is approximately~$\SI{992}{\second}$. In this case, we choose~$\Delta t = \SI{1000}{\second}$. The values of~$\Delta t$ for each dataset shown in Sec.~\ref{Sec:RESULTS} are reported in Table~\ref{Tab:Experimental:Parameters}.

\begin{figure}[b!]
	\centering
	\includegraphics{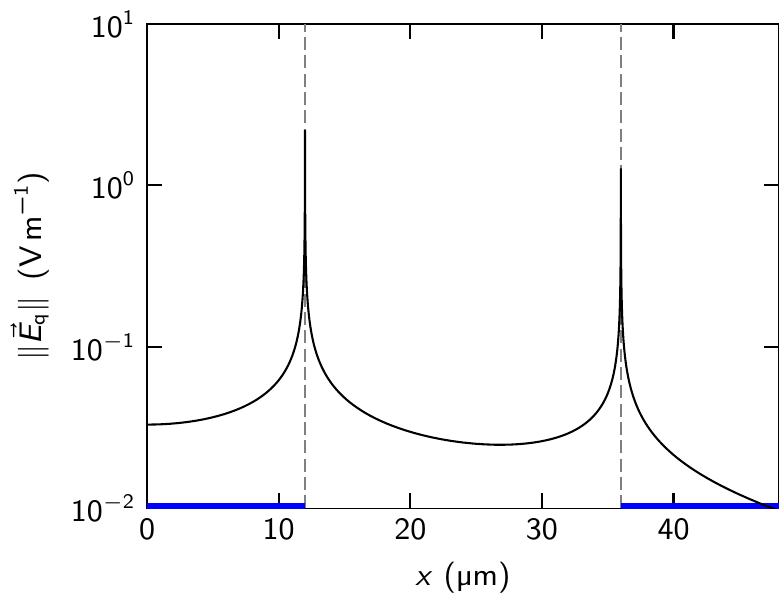}
	\caption{Qubit electric field~$\norm*{\vec{E}_{\text{q}}}$ for~$\phi_0 = \SI{1}{\volt}$ vs.~width~$x$ at one value of the height~$z = \SI{1.5}{\nano\meter}$. The origin of the graph is at~$x = 0$, corresponding to the middle point of the strip. Due to the symmetry of the CPW segment with respect to its longitudinal axis (i.e., the $y$-axis; not shown), we display~$\norm*{\vec{E}_{\text{q}} (x)}$ only for half of the CPW segment, for~$x > 0$. The extent of the conducting sections of the CPW is indicated by the thick blue lines. The dashed black vertical lines are placed at the edge of each conductor; the left line corresponds to the edge of the strip and the right line to the edge of the ground plane. \label{Fig:APPENDIX:Electric:Field}}
\end{figure}

\section{QUBIT ELECTRIC FIELD}
	\label{App:QUBIT:ELECTRIC:FIELD}

As explained in App.~\ref{App:DEVICE:AND:SETUP}, the qubit capacitor is a Greek cross formed by two CPW strips of length~$L$. Since~$L \gg S+W$, we approximate the qubit capacitor as a CPW segment of infinite length; we additionally assume that the capacitor is made of an infinitesimally thin conducting sheet. When determining~$\vec{E}_{\text{q}}$, we can thus restrict ourselves to points within the CPW vertical cross section.

We determine~$\vec{E}_{\text{q}}$ by means of a conformal mapping technique. A conformal map is a function that locally preserves angles, allowing us to transform the CPW geometry into that of a much simpler infinite parallel-plate capacitor; the map function is given by Eq.~(25) in the work of Ref.~\cite{Murray:2018}. We then use this map to transform the electric field of the parallel-plate capacitor into that of the CPW. The electric field is proportional to the qubit electric potential with respect to ground, or zero-point voltage; given the qubit plasma frequency~$f_{\text{p}} = \sqrt{8 E_{\text{J}} E_{\text{c}}} / h$, the zero-point voltage reads
\begin{equation}
\phi_0 \simeq \sqrt{\dfrac{h f_{\text{p}}}{2 C_{\text{q}}}} = \dfrac{e}{C_{\text{q}}} \left( \dfrac{E_{\text{J}}}{2 E_{\text{c}}} \right)^{\!\! 1/4} \sim \SI{4}{\micro\volt} .
\end{equation}

In order to generate~$g$, we evaluate~$\norm*{\vec{E}_{\text{q}}}$ at randomly picked points~$(x,z)$ corresponding to Q-TLS positions. These points are confined within the cross-section region introduced above. The cross section is centered on the middle point of the strip and has a length of~\SI{96}{\micro\meter} and a height of~\SI{3}{\nano\meter}; the left and right edges of the cross section extend~\SI{12}{\micro\meter} into the ground plane and the top edge corresponds to the oxide layer's top edge. Figure~\ref{Fig:APPENDIX:Electric:Field} shows~$\norm*{\vec{E}_{\text{q}} (x,z)}$.

\bibliography{bibliography}

\end{document}